\documentclass[aps,pra,superscriptaddress,amsmath,amssymb,preprintnumbers,floatfix,showpacs,12pt]{revtex4}
\usepackage{amssymb}
\usepackage{epsfig}

\begin{document}
\title{ Revival-collapse phenomenon in the quadrature squeezing of the
multiphoton Jaynes-Cummings model with the binomial states}
\author{ Faisal A. A. El-Orany }
\email{el_orany@yahoo.com}
 \affiliation{Department of Mathematics  and computer Science,
Faculty of Science, Suez Canal University 41522,
 Ismailia, Egypt}

\date{\today}

\begin{abstract}
In this paper we study the interaction between  two-level atom and
 quantized  single-mode field, namely, Jaynes-Cummings
model (JCM). The field and the atom are initially prepared in the
binomial state and the excited atomic state, respectively. For this
system we prove that the revival-collapse phenomenon exhibited  in
the atomic inversion of the standard  JCM can be numerically
(naturally) manifested in the evolution of the squeezing factor of
the three-photon (standard) JCM provided that the initial
photon-number distribution of the radiation has a smooth envelope.

\end{abstract}

  \maketitle

\section{Introduction}
Jaynes-Cummings model (JCM) \cite{jay1} is one of the fundamental
systems in the quantum optics. The simplest form of the JCM is the
single quantized mode interacting with the two-level
  atom. Various phenomena have been realized for this system such as revival-collapse
  phenomenon (RCP) in the evolution of the atomic inversion \cite{eber},
  sub-Poissonian statistics   and squeezing, e.g., \cite{bruce}.
Actually, the RCP represents  the most important phenomena reported
to this model since it manifests  the granular nature of the initial
field distribution \cite{zaber1} as well as  the strong entanglement
between the radiation field and the atom. The RCP has been observed
via the one-atom mazer \cite{remp} and also using technique similar
to that of the NMR refocusing \cite{meu}.
 It is worth mentioning that the RCP has been
also remarked in the evolution of different quantities in the
nonlinear optics such as the mean-photon number of the Kerr
nonlinear coupler \cite{peri} and
 the photon-number distribution  of
the single-mode \cite{dutta} and two-mode \cite{mary} squeezed
coherent states with complex squeeze and displacement parameters. In
the latter two cases the RCP occurs in the photon-number domain
rather than in the interaction time domain.

  There is an important class of nonclassical states, namely,
binomial states. The binomial state (BS) is an intermediate state
between the Fock state and the coherent state \cite{salh}, i.e. it
is  linear combination of Fock states weighted by binomial
distributions.
 The BS can exhibit many nonclassical effects, e.g., squeezing,
sub-Poissonian statistics and negative values in the Wigner function
\cite{valo}.
 Recently, the BSs have
been proposed as reference field states in schemes  measuring the
canonical phase of quantum electromagnetic fields \cite{peg}. The BS
can be generated by a classical current interacting with two
quantized radiation fields \cite{salh,dato} as well as via quantum
state engineering \cite{wu}. In the latter technique the resonant
interaction of $N'$ consecutive two-level atoms with the cavity
initially prepared in its vacuum state is constructed and the
desired cavity field state can be obtained from total state
reduction by performing measurement on the atoms coming out of the
cavity \cite{vogel}. Quite recently, an efficient scheme for
generating and detecting  two-photon  generalized binomial state in
a single-mode high-$Q$ cavity is  described  in \cite{franc}.
 Evolution of the BS
with the JCM has been investigated for the single-photon JCM
\cite{sing}, the two-photon JCM \cite{two} and the single-photon
 Kerr-nonlinear JCM \cite{kerrr}. The object of these
studies is to investigate the construction of  different phenomena
 when the field evolves gradually from the Fock
state to the coherent state. It is worth referring that the
evolution of the BS with the JCM can provide different behaviors
than those with the coherent state, e.g., under certain conditions
the evolution of the atomic inversion related to the BS exhibits a
very steady beat phenomenon similar to that found in the classical
physics \cite{kerrr}. The superposition of the BS (SBS) has been
developed
 \cite{abda,binm} as:
\begin{equation}
|M,\eta\rangle_{\epsilon}=\sum_{n=0}^{M}C_n^M(\eta,\epsilon)
|n\rangle, \label{bin1}
\end{equation}
where the  coefficient $C_n^M(\eta,\epsilon)$ takes the form
\begin{eqnarray}
\begin{array}{lr}
 C_n^M(\eta,\epsilon)
=\lambda_{\epsilon}\sqrt{\frac{M!}{(M-n)!n!}}
\eta^n(1-|\eta|^2)^{[\frac{M-n}{2}]}[1+(-1)^n\epsilon],\\
\\
|\lambda_{\epsilon}|^{-2}=1+\epsilon^2+2(1-2|\eta|^2)^M\epsilon,
\label{bin2}
\end{array}
\end{eqnarray}
where $M$ is a positive integer, $0<|\eta|\leq 1$ and $\epsilon$ is
a parameter taking one of the values $0,1$ and $-1$ corresponding to
BS, even BS and odd BS, respectively.  Throughout the investigation
we consider $\eta$ to be real. In the limiting cases
$(\epsilon,\eta)\rightarrow (0,1)$ and $(\eta,M)\rightarrow
(0,\infty)$ such that $M\eta^2=\alpha^2$ the state (\ref{bin1})
reduces to the Fock state $|M\rangle$ and the superposition of the
coherent state $|\alpha\rangle_{\epsilon}$ \cite{schr},
respectively.

There is another type of the SBS, which is called the phased
generalized binomial state \cite{binm}. This type of state is
represented by  the superposition of the even or odd  binomial
states. As an example we give   the definition of the
orthogonal-even binomial state as

\begin{equation}
|M,\eta\rangle_{e}=A\sum_{n=0}^{[M/4]}C_{4n}^M(\eta,0) |4n\rangle,
\label{bin1b}
\end{equation}
where $C_{4n}^M(\eta,0)$ can be obtained from (\ref{bin2}) and $A$
is the normalization constant having the form
\begin{equation}
A^2=\frac{4}{1+(1-2|\eta|^2)^M+2{\rm Re}(1-|\eta|^2+i|\eta|^2)^M}
. \label{bin2b}
\end{equation}
Using  appropriate  limit  the state (\ref{bin1b}) tends to the
orthogonal-even coherent state \cite{lyn}. It is worth mentioning
that the common property of the binomial states is that   the
probability of detecting $m$ quanta when $m>M$ is zero.  From the
above information one can realize that the SBS is one of the most
generalized states in quantum optics.

Recently, for the
 JCM it has been shown that  there is  a relationship between
the atomic inversion and the quadrature squeezing  \cite{fas1,fas2}.
 More illustratively, for particular type of initial states, e.g.
$l$-photon coherent states,
  the squeezing factors  can naturally
 provide complete information on the corresponding atomic inversion.
 Nevertheless, for the initial
 coherent state it has been numerically shown that
the evolution of the quadrature squeezing of the three-photon JCM
reflects the RCP involved in the atomic inversion of the standard,
i.e. the single-photon, JCM. These relations have been obtained
based on the fact that  for the initial $l$-photon coherent state
and coherent state the harmonic approximation is applicable.
 In this paper we show that
these relations  exist also for any arbitrary initial field states
provided that their photon-number distributions have smooth
envelopes. In doing so we study the evolution of the JCM with the
SBS. For this system we obtain various interesting results. For
instance, we show that the relations between the atomic inversion
and the quadrature squeezing are sensitive to the interference in
phase space. Additionally,  the odd $N$th-order squeezing of the
standard JCM with the even-orthogonal  binomial state exhibits  RCP
as that of the corresponding atomic inversion. The motivation of
these relations is that the RCP exhibited in the evolution of the
atomic inversion can be measured by the homodyne detectors
\cite{homo}. This is supported by the recent developments in the
cavity QED in which the homodyne detector technique has been applied
to the single Rydberg atom and one-photon field for studying the
field-phase evolution of the  JCM \cite{haroc}.

We construct  the paper   in the following order: In section 2 we
give
 the basic relations and equations related to the system under
consideration.
 In sections 3 and 4 we investigate naturally and numerically the occurrence of
the RCP in the higher-order squeezing. In section 5 we summarize the
main  results.

\section{Basic equations and relations}
In this section we give the basic relations and equations, which
will be used in the paper. Specifically, we develop
 the Hamiltonian of the system and its wavefunction  as well as  the definition of the quadrature
squeezing. Also we shed  light on the relation between the
photon-number distribution and the atomic inversion.

The Hamiltonian controlling the interaction between the two-level
atom and the $k$th-photon single-mode field in the rotating wave
approximation
 is \cite{rose}:
\begin{equation}
\frac{\hat{H}}{\hbar}=\omega_{0}\hat{a}^{\dagger}\hat{a}+\frac{1}{2}
\omega_{a}\hat{\sigma}_{z}+\lambda (\hat{a}^{k}\hat{\sigma}_{+} +
\hat{a}^{\dagger k}\hat{\sigma}_{-}),
 \label{6}
\end{equation}
where $\hat{\sigma}_{\pm}$ and $\hat{\sigma}_{z}$ are the Pauli
spin operators; $\hat{a}\quad (\hat{a}^{\dagger})$ is the
annihilation (creation) operator denoting  the cavity mode,
$\omega_{0}$ and $\omega_{a}$ are the frequencies of the cavity
mode and the atomic transition, respectively; $\lambda$ is the
atom-field coupling constant and $k$ is the transition parameter.

\begin{figure}
  \vspace{0cm}
    \includegraphics[width=0.86\linewidth]{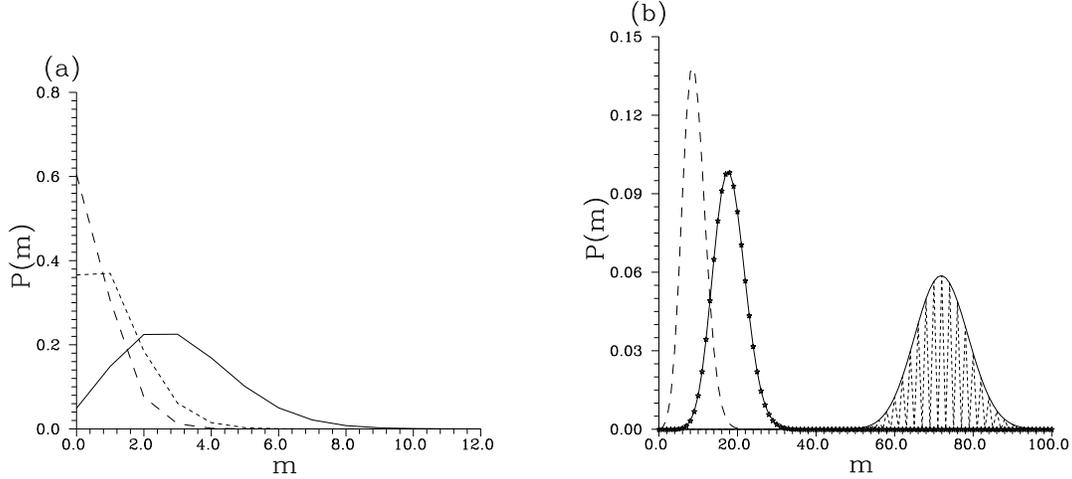}
\caption{ The $P(m)$ against $m$ for the BS. (a) $\eta=0.1$ and
$M=50$ (long-dashed curve), $100$ (short-dashed curve) and $370$
(solid curve). (b) $(\eta, M)=(0.3,100)$ (long-dashed curve),
$(0.3,200)$ (star-centered curve) and $(0.6,200)$ (solid curve). The
short-dashed curve in (b) is given for $(\epsilon, \eta,
M)=(1,0.6,200)$.
 }
\end{figure}

\begin{figure}
  \vspace{0cm}
    \includegraphics[width=0.86\linewidth]{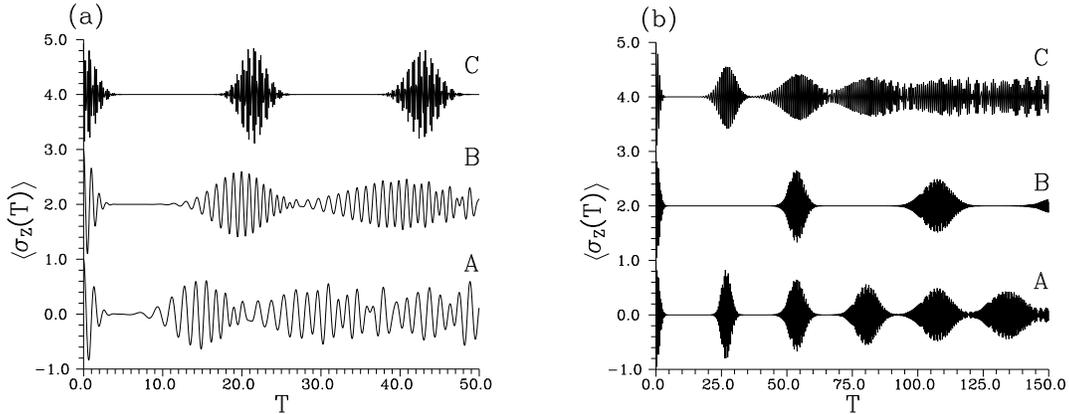}
\caption{  The atomic inversion $\langle\sigma_{z}(T)\rangle$
against the scaled time $T$ for $k=1$ when the field is initially
prepared in the SBS with different values of $\eta, M $ and
$\epsilon$.
 (a) $(\eta,M,\epsilon)=(0.1,370,0)$
(curve A) and $(0.3,100,0)$ (curve B), whereas the curve C is given
for  the initial orthogonal-even binomial state with
$(M,\eta)=(370,0.7)$. (b) $M=200$ and $(\eta,\epsilon)=(0.6,1)$
(curve A), $(0.6,0)$ (curve B) and $(0.3,0)$ (curve C). The curves
are shifted from the bottom by $0,2,4$.
 }
\end{figure}

We consider that the field and atom are initially prepared in the
SBS (\ref{bin1}) and
 the excited atomic state $|+\rangle$, respectively. Also we
 restrict the investigation to
 the exact resonance case.
Under these conditions the dynamical state  of the system can be
expressed as
\begin{equation}
|\Psi(T)\rangle=
\sum\limits_{n=0}^{M}C_n^M(\eta,\epsilon)\left[\cos(T\nu_{n,k})|
+,n\rangle-i\sin(T\nu_{n,k})|-,n+k\rangle
\right], \label{10}
\end{equation}
where $T=\lambda t, \nu_{n,k}=\sqrt{\frac{(n+k)!}{n!}}$ and
$|-\rangle$ denotes the ground atomic state. The atomic inversion
associated with (\ref{10}) is
\begin{equation}
\langle\sigma_{z}(T)\rangle=\sum\limits_{n=0}^{M}P(n)\cos(2T\nu_{n,k}),
\label{13}
\end{equation}
where $P(n)=|C_n^M(\eta,\epsilon)|^2$. To understand the relation
between $P(m),  \langle\sigma_{z}(T)\rangle$ and  quadrature
squeezing we  plot $P(m)$ and $\langle\sigma_{z}(T)\rangle$ in Figs.
1 and 2, respectively, for the given values of the parameters. Fig.
1(a) gives the development of the binomial state to the coherent
state. This is obvious from the solid curve in Fig. 1(a) as well as
the curve A in the Fig. 2(a), which represents the RCP of the
coherent state with $\alpha=\sqrt{M\eta^2}=\sqrt{3.7}$ (we have
checked this fact). Generally, the comparison between the
corresponding curves in Figs. 1 and  2 shows
 when $P(m)$ exhibits
smooth envelope the $\langle\sigma_{z}(T)\rangle$ provides the RCP.
Moreover, the interference in phase space manifests itself as two
times revival patterns in the evolution of the
$\langle\sigma_{z}(T)\rangle$ compared to those related to the BS
(compare the curves A and B in Fig. 2(b)). Also from the dashed
curve in Fig. 1(b)  the maximum value of the $P(m)$ is close to
$m\simeq \bar{n}=\langle\hat{a}^{\dagger}(0)\hat{a}(0)\rangle$.
 For the future purpose we
have plotted curve C in Fig. 2(a) for the
$\langle\sigma_{z}(T)\rangle$ of the initial orthogonal-even BS.

The different moments of the operators $\hat{a}^{\dagger}$ and
$\hat{a}$ for the state (\ref{10}) can be evaluated as
\begin{eqnarray}
\begin{array}{lr} \langle\hat{a}^{\dagger s_{2}}(T)
\hat{a}^{s_{1}}(T)
\rangle=\sum\limits_{n=0}^{M-s_1}\left(C_{n+s_2}^{M}(\eta,\epsilon)\right)^{*}\
C_{n+s_1}^M(\eta,\epsilon) \left[
\cos(T\nu_{n+s_2,k})\cos(T\nu_{n+s_1,k})
\frac{\sqrt{(n+s_{1})!(n+s_{2})! }}{n!}\right.\\
\\
+\left. \sin(T\nu_{n+s_2,k})\sin(T\nu_{n+s_1,k})
\frac{\sqrt{(n+k+s_{1})!(n+k+s_{2})! }}{(n+k)!} \right],
\label{12}
\end{array}
\end{eqnarray}
where $s_{1}$ and $s_{2}$ are positive integers and $M<s_1$.
Finally, the $N$th-order quadrature squeezing operators are defined
by $\hat{X}_N=\frac{1}{2}(\hat{a}^{N}+\hat{a}^{\dagger N}),\quad
\hat{Y}_N=\frac{1}{2i}(\hat{a}^{N}-\hat{a}^{\dagger N})$, where $N$
is a positive integer. The squeezing factors associated with the
$\hat{X}_N$ and $\hat{Y}_N$ can be, respectively, expressed as
\cite{hig}:

\begin{eqnarray}
\begin{array}{lr}
F_N(T) = \langle\hat{a}^{\dagger N}(T)\hat{a}^{N}(T)\rangle+{\rm
Re} \langle\hat{a}^{2N}(T)\rangle-2({\rm
Re}\langle\hat{a}^{N}(T)\rangle)^{2},
\\
\\
S_N(T)=
 \langle\hat{a}^{\dagger N}(T)\hat{a}^{N}(T)\rangle
-{\rm Re}\langle\hat{a}^{2N}(T)\rangle-2({\rm
Im}\langle\hat{a}^{N}(T)\rangle)^{2}. \label{14}
\end{array}
\end{eqnarray}

Now we are in a position to investigate the relation between the
atomic inversion of the standard, i.e. $k=1$, JCM denoting by
$\langle\sigma_{z}(T)\rangle_{k=1}$ and the quadrature squeezing.
This will be done in the following sections.

\section{Natural approach}
Natural approach is based on the fact: the quantity
$\langle\sigma_{z}(T)\rangle+
\langle\hat{a}^{\dagger}(T)\hat{a}(T)\rangle$ is a constant of
motion and hence  $\langle\sigma_{z}(T)\rangle$ and
$\langle\hat{a}^{\dagger}(T)\hat{a}(T)\rangle$ can carry information
on each others \cite{fas1}. Furthermore, this approach can be
generalized to find a relation between $\langle\sigma_{z}(T)\rangle$
and $\langle\hat{a}^{\dagger N}(T)\hat{a}^{N}(T)\rangle$, where $N$
is a positive integer, as we show shortly. The discussion will be
restricted to the case $k=1$.
 Now the question is that for the JCM  which type of the binomial
states making the $N$th-order squeezing factors provide complete
 information on the corresponding atomic inversion? The
answer to this question
  can be realized from (\ref{14}). Precisely,
when there is a type of binomial states satisfying simultaneously
the conditions:

\begin{equation}
\langle\hat{a}^{N}(T)\rangle=0,\quad
\langle\hat{a}^{2N}(T)\rangle=0. \label{extr1}
\end{equation}
In this case  the squeezing factors reduce to
$\langle\hat{a}^{\dagger N}(T)\hat{a}^{N}(T)\rangle$, which can be
connected with the corresponding $\langle\sigma_{z}(T)\rangle$.
 As an example of these states  is
the orthogonal-even binomial states (\ref{bin1b}) provided that the
squeezing order $N$ is odd integer, i.e. $2N+1$. For this case one
can easily check that  the conditions (\ref{extr1}) are fulfilled
and hence the squeezing factors reduce to

\begin{eqnarray}
\begin{array}{rl}
F_{2N+1}(T)=\langle\hat{a}^{\dagger
2N+1}(T)\hat{a}^{2N+1}(T)\rangle= \langle\hat{a}^{\dagger
2N+1}(0)\hat{a}^{2N+1}(0)\rangle+
(N+\frac{1}{2})\langle\hat{a}^{\dagger 2N}(0)\hat{a}^{2N}(0)\rangle
\\\\
-(N+\frac{1}{2}) A^2
\sum\limits_{n=0}^{[M/4]}|C_{4n}^{M}|^2\frac{(4n)!}{(4n-2N)!}\cos(2T\nu_{4n,1}).
 \label{extr2}
\end{array}
\end{eqnarray}
Using suitable  limits for the summation in (\ref{extr2}) and by
means of the following relation
\begin{equation}
M!=(M-2N)!M^{2N}\prod\limits_{j=0}^{2N-1}(1-\frac{j}{M})
\end{equation}
we arrive at

\begin{eqnarray}
\begin{array}{rl}
F_{2N+1}(T)= \langle\hat{a}^{\dagger
2N+1}(0)\hat{a}^{2N+1}(0)\rangle+
(N+\frac{1}{2})\langle\hat{a}^{\dagger 2N}(0)\hat{a}^{2N}(0)\rangle
\\\\
-(N+\frac{1}{2}) |\eta|^{4N}M^{2N}\left[
\prod\limits_{j=0}^{2N-1}(1-\frac{j}{M})\right]A^2
\sum\limits_{n=0}^{[M/4-2N]}|C_{4n}^{M-2N}|^2\cos(2T\nu_{4n+2N,1}).
 \label{extr222}
\end{array}
\end{eqnarray}
For finite (large) values of $N$ ($M$) with $0<\eta<1$, i.e. the
$P(m)$ has smooth envelope, we can use the substitutions
 $\nu_{4n+2N,1}\simeq \nu_{4n,1}$ and $|C_{4n}^{M-2N}|^2\simeq |C_{4n}^{M}|^2$
 and hence the
expression (\ref{extr222}) can be modified to give the rescaled
squeezing factor
 $W_N(T)(=\langle\sigma_{z}(T)\rangle_{k=1})$ through the
 relation:
\begin{equation}
 W_{N}(T)=\frac{ 2\langle\hat{a}^{\dagger
2N+1}(0)\hat{a}^{2N+1}(0)\rangle+ (2N+1)\langle\hat{a}^{\dagger
2N}(0)\hat{a}^{2N}(0)\rangle-2F_{2N+1}(T) }
{(2N+1)\langle\hat{a}^{\dagger 2N}(0)\hat{a}^{2N}(0)\rangle_b},
\label{extr31}
\end{equation}
where the subscript $b$ in the denominator means that the quantity
$\langle\hat{a}^{\dagger 2N}(0)\hat{a}^{2N}(0)\rangle$ is related to
the BS. Now we are in a position to check the validity of the
(\ref{extr31}). Thus we plot (\ref{extr31}) in  Fig. 3 for  the
third-order squeezing and the
 given values of the interaction parameters. We should stress that
 in Fig. 3 we have used the explicit form for $F_{2N+1}$ given by
(\ref{extr2}). The comparison between the curve C in Fig. 2(a) and
Fig. 3 demonstrates our conclusion:
 for particular type of  binomial
states the squeezing factor can provide complete information on the
corresponding atomic inversion. The origin in this is that the
expressions of the  $\langle\sigma_{z}(T)\rangle$ and
$\langle\hat{a}^{\dagger N}(T)\hat{a}^{N}(T)\rangle$ depend on the
diagonal elements of the density matrix of the system under
consideration.

\begin{figure}
  \vspace{0cm}
    \includegraphics[width=0.86\linewidth]{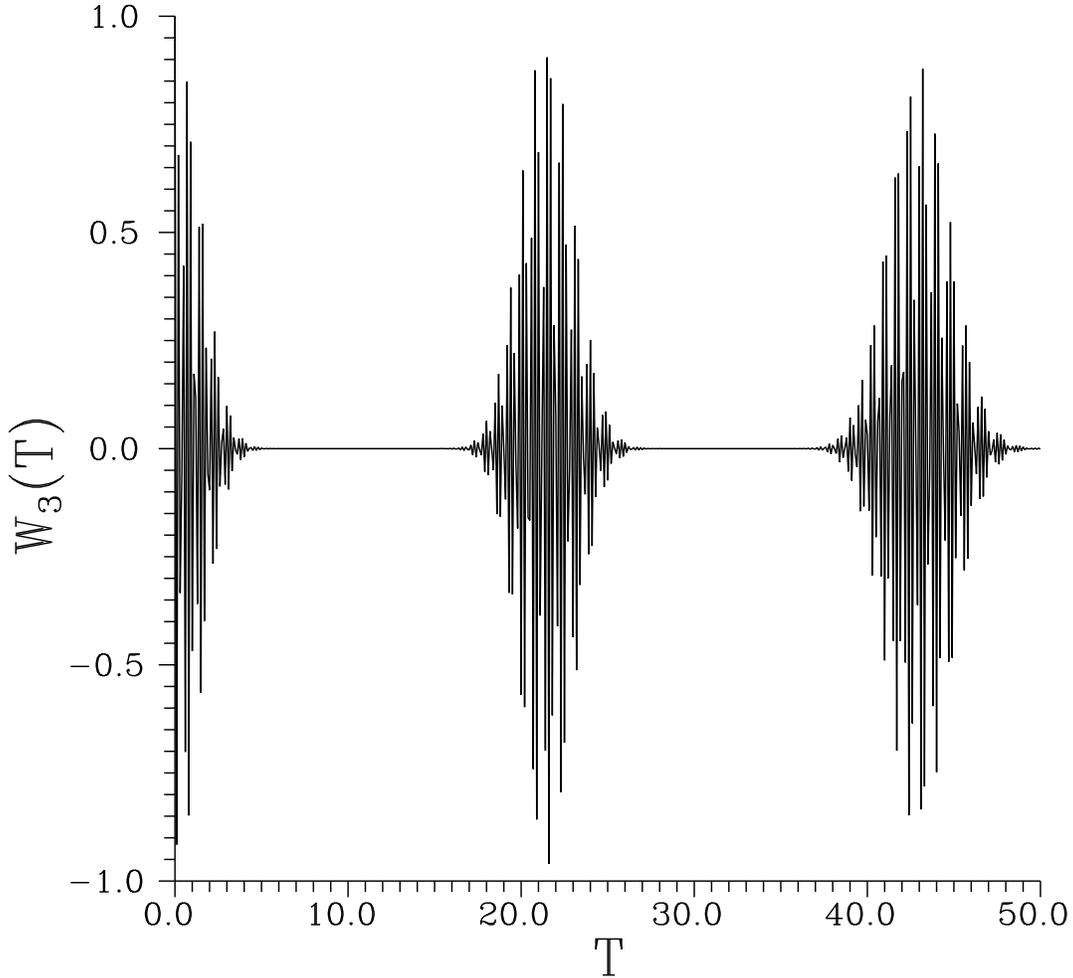}
\caption{  The third-order rescaled squeezing factor
 given by (14) against the scaled time $T$, when the field is initially
 in the orthogonal-even
 BS with $(M,\eta)=(370,0.7)$.
 }
\end{figure}

\section{Numerical  approach}
In this section we study the possibility of obtaining information on
the $\langle\sigma_{z}(T)\rangle_{k=1}$ from  the squeezing factors
of the $k$th-photon JCM, when the field is initially prepared in the
SBS. Our object is to find the value of the transition parameter
$k\quad (k>2)$ for which one or both of the squeezing factors
produce RCP as that involved in the
$\langle\sigma_{z}(T)\rangle_{k=1}$ \cite{fas1}. From (\ref{14}) the
RCP can likely occur in the $F_N(T)\quad (S_N(T))$ only when
 $ {\rm Re}\langle\hat{a}^{N}(T)\rangle=0\quad ({\rm
Im}\langle\hat{a}^{N}(T)\rangle=0)$ since these quantities are
squared, i.e. they destroy the RCP if it exists. According to this
fact the occurrence of the RCP in $F_N$ or in $S_N$ depends on the
values of the $\epsilon$ and $N$. Moreover, for $k>2$ the quantity
$\langle\hat{a}^{\dagger N}(T)\hat{a}^N(T)\rangle$ exhibits chaotic
behavior and hence we can use $\langle\hat{a}^{\dagger
N}(T)\hat{a}^N(T)\rangle\simeq \langle\hat{a}^{\dagger
N}(0)\hat{a}^N(0)\rangle$. From this discussion we can conclude that
if the squeezing factors exhibit RCP this will be related to
 the quantity $ {\rm Re}\langle\hat{a}^{2N}(T)\rangle$.
Thus we treat this quantity in a greater details. From (\ref{12})
and after minor algebra we arrive at

\begin{eqnarray}
\begin{array}{lr}
\langle\hat{a}^{2N}(T) \rangle=\frac{|\eta|^{2N}M^N}
{(1-|\eta|^2)^N} \sum\limits_{n=0}^{M}|C_{n}^{M}(\eta,\epsilon)|^2
\left\{\sqrt{\prod\limits_{j=1}^{2N}(1-\frac{(n+2N-j)}{M})}
\right\}\\
\\
\times \Bigl[ \cos(T\nu_{n+2N,k})\cos(T\nu_{n,k}) +
\sqrt{\prod\limits_{j=0}^{2N-1}\frac{(1+\frac{(k+2N-j)}{n})}{(1+\frac{2N-j}{n})}}
\sin(T\nu_{n+2N,k})\sin(T\nu_{n,k})
 \Bigr].
\label{m112}
\end{array}
\end{eqnarray}
In  (\ref{m112}) we have extended the upper limit of the summation
from $M-2N$ to $M$ using the fact $l!=-\infty$ when $l<0$
\cite{moya} because our goal is to compare this expression with that
of the $\langle\sigma_{z}(T)\rangle_{k=1}$. Moreover, we assume that
$M>>2N$, $0<\eta< 1$ and $\bar{n}$ is very large. Therefore, the
quantity in the square root in the second line of (\ref{m112}) tends
to unity. Additionally, for $\epsilon=0,i$ the $P(n)$ exhibits
smooth envelope and then the terms contributing effectively to the
summation in (\ref{m112}) are those close to $n\simeq
\bar{n}=M|\eta|^2$. In this case the quantity
 in the curely curves in (\ref{m112}) can be simplified as

\begin{equation}
\sqrt{\prod\limits_{j=1}^{2N}(1-\frac{(n+2N-j)}{M})} =
\sqrt{\prod\limits_{j=1}^{2N}(1-|\eta|^2-\frac{(2N-j)}{M}) } \simeq
\left(1-|\eta|^2\right)^N,
 \label{1mm}
\end{equation}
where we have considered $\vartheta/M\rightarrow 0$ since
$\vartheta$ is a finite c-number and $M>>\vartheta$. On the other
hand, when $\epsilon=1$, say, the $P(n)$
 exhibits oscillatory behavior with  maximum value  around
 $n\simeq \bar{n}$ (see the dashed curve in the Fig. 1(b)) and we arrive at

\begin{equation}
\sqrt{\prod\limits_{j=1}^{2N}(1-\frac{(n+2N-j)}{M})} \simeq
\left(1-\frac{\bar{n}}{M}\right)^N=
\left(1-|\eta|^2\frac{(1-z^{M-1})}{1+z^M}\right)^N
=(1-|\eta|^2)^N\left(\frac{1+z^{M-1}}{1+z^M}\right)^N, \label{3mm}
\end{equation}
where we have used the mean-photon number of the even binomial
states as \cite{abda}:
\begin{equation}
\bar{n}=|\eta|^2M\frac{(1-z^{M-1})}{(1-z^{M})},\qquad z=1-2|\eta|^2.
\label{73mm}
\end{equation}
 It is evident that $|z|<1$ for $0<\eta<1$ and then $z^{M-1}\simeq 0$ where $M$ is very
 large.
 Thus the result given by (\ref{1mm}) is valid for all values of
$\epsilon$ and  hence the expression (\ref{m112}) reduces to

\begin{figure}
  \vspace{0cm}
    \includegraphics[width=0.86\linewidth]{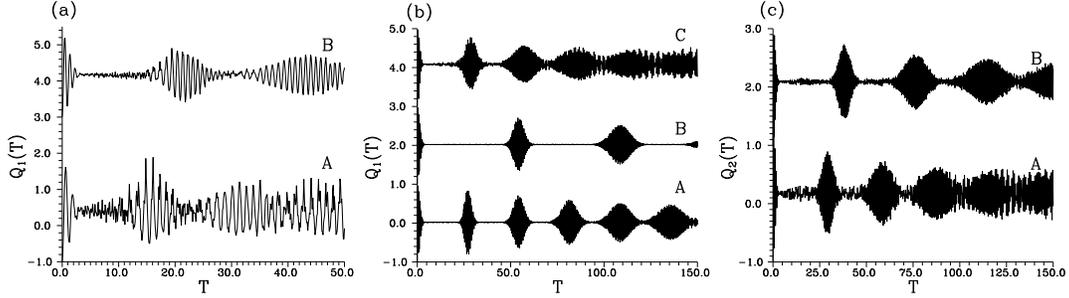}
\caption{  The rescaled squeezing factors $Q_1(T)$ and $Q_2(T)$ as
indicated against the scaled time $T$ for different values of
$\eta,M$ and $\epsilon$.
 (a) $(\eta,M,\epsilon)=(0.1,370,0)$ (curve A) and $(0.3,100,0)$ (curve
B). (b) $M=200$ and $(\eta,\epsilon)=(0.6,1)$ (curve A), $(0.6,0)$
(curve B) and $(0.3,0)$ (curve C). (c)
$(\eta,M,\epsilon)=(0.3,200,0)$ (curve A) and $(0.3,370,0)$ (curve
B).  The curves in (b) and (c) are shifted from the bottom by
$0,2,4$ and $0,2$, respectively, whereas in (a) are shifted by
$0,4$.
 }
\end{figure}

\begin{equation}
\langle\hat{a}^{2N}(T) \rangle\simeq |\eta|^{2N}M^N
\sum\limits_{n=0}^{M}|C_{n}^{M}(\eta,\epsilon)|^2
\cos[T(\nu_{n+2N,k}-\nu_{n,k})]. \label{m13}
\end{equation}
Comparison between (\ref{13}) (i.e.
$\langle\sigma_{z}(T)\rangle_{k=1}$) and (\ref{m13})   shows
 that the two expressions can  provide similar dynamical
behavior only when the arguments of the cosines  are comparable.
This is regardless of the different scales resulting from the
pre-factor $M|\eta|^{2N}$ in (\ref{m13}). The proportionality factor
$\mu_N$, say, which makes the dynamical behaviors in the two
expressions similar, can be evaluated from the following relation
\begin{eqnarray}
\begin{array}{lr}
\mu_N=\frac{\nu_{n+2N,k}-\nu_{n,k}}{2\sqrt{n+1}},\\
\\
=\frac{n^{\frac{k}{2}}\sqrt{\prod\limits_{j=1}^{k}(1+\frac{j}{n})}\left[
\prod\limits_{j=1}^{2N}(n+k+j) -\prod\limits_{j=1}^{2N}(n+j)\right]
}
{2n^{2N+\frac{1}{2}}\sqrt{1+\frac{1}{n}}\sqrt{\prod\limits_{j=1}^{2N}(1+\frac{j}{n})}
\left[\sqrt{\prod\limits_{j=1}^{2N}(1+\frac{k+j}{n})}
+\sqrt{\prod\limits_{j=1}^{2N}(1+\frac{j}{n})}\right] } .
\label{m155}
\end{array}
\end{eqnarray}
It is worth recalling that  $\bar{n}$ is very large, $P(n)$ exhibits
smooth envelope, i.e. $n\simeq \bar{n}$, and the squeezing-order $N$
is finite. Therefore, by applying the Taylor expansion for different
square roots in (\ref{m155}) we obtain \cite{fas2}:

\begin{equation}
\mu_N\simeq \frac{1}{4} [
2Nk\bar{n}^{\frac{k-3}{2}}+\bar{n}^{\frac{k-5}{2}}(...)+
\bar{n}^{\frac{k-7}{2}}(...)+...]. \label{ol7o}
\end{equation}
From (\ref{ol7o}) it is evident that the RCP can occur in the
squeezing factor only when $k=3$ and hence $\mu_N=\frac{3N}{2}$. In
this case we have neglected such type of  terms
$\bar{n}^{-1},\bar{n}^{-2},..., $ where $\bar{n}$ is very large.
 From the above investigation one can realize  that  the $N$th-order rescaled squeezing factor,
which can give complete information on the
$\langle\sigma_{z}(T)\rangle_{k=1}$, is
\begin{equation}
Q_{N}(T)=\frac{\langle\hat{n}(0)\rangle_b^N -V_{N}(T)}
{\langle\hat{n}(0)\rangle_b^N}, \label{29qqq}
\end{equation}
where
\begin{equation}    
V_{N}(T)  = \left\{
\begin{array}{lr}
S_{N}(\frac{3T}{2N})
\;\;&{\rm for}\;\epsilon=0 ,\\
\\
F_{N}(\frac{3T}{2N})
\;\;&{\rm for}\;\epsilon=i ,\\
\\
S_{N}(\frac{3T}{2N})=F_{N}(\frac{3T}{2N}) \;\;&{\rm
for}\;\epsilon=\pm 1,\;\; N=2m'+1,\\
\\
S_{N}(\frac{3T}{2N}) \;\;&{\rm for}\;\epsilon=\pm 1,\;\; N=2m'
\end{array} \right. \label{8a}
\end{equation}
and $m'$ is a positive integer. In the derivation of the formula
(\ref{29qqq}) we have considered that the mean-photon numbers of the
BS and the SBS are the same. This  is correct for $0<\eta<1$ and
large $\bar{n}$.
 It is worth mentioning that  the formula  (\ref{29qqq})
is valid for the initial superposition of the coherent states,
too. Now we  check the validity of  (\ref{29qqq}) by  plotting
 Figs. 4(a)--(c) for the given values of the interaction parameters.
The comparison between the curves in Figs. 4(a)-(b) and the
corresponding ones in the Figs. 2 leads to the following fact:
 when $0<\eta<1$ and  $\bar{n}$
is large, regardless of the values of $\epsilon$, the $Q_1(T)$
copies well with the $\langle\sigma_{z}(T)\rangle_{k=1}$.
Nevertheless, when $\bar{n}$ is relatively small (with $P(m)$ has a
smooth envelope)
 the RCP can be established in $Q_1(T)$, but
the overall behavior could be  different from that of the
 $\langle\sigma_{z}(T)\rangle_{k=1}$. This result is
obvious when we compare the curves A in Fig. 2(a) and  Fig. 4(a),
where one can observe  $|Q_1(T)|>1$.
  Fig. 4(c) is given for
the higher-order squeezing. The comparison between the curve A in
this figure and the curve C in Fig. 4(b) leads to that the normal
squeezing can provide better information on the
$\langle\sigma_{z}(T)\rangle_{k=1}$ than the amplitude-squared
squeezing.  Nevertheless, the information obtained from the
higher-order squeezing can  be improved by increasing the value of
the $\bar{n}$ (compare the curves A and B in Fig. 4(c)).

\section{Conclusion }
In this paper we have shown that for the JCM there is a relationship
between the quadrature squeezing  and the atomic inversion provided
that the initial photon-number distribution exhibits smooth
envelope.
 This fact has been proved using one of the most general
quantum state, namely, the superposition of the binomial states.
Precisely, we have shown that for particular types of the initial
binomial states the $N$th-order squeezing factor can naturally give
complete information on the corresponding atomic inversion. Also we
have numerically shown that the $N$th-order squeezing factor of the
three-photon JCM can provide complete information on the
$\langle\sigma_{z}(T)\rangle_{k=1}$. These relations exist only when
 the $P(m)$ exhibits smooth envelope and $\bar{n}$ is large.
Moreover, as the squeezing order $N$ increases the values of the
$\bar{n}$ have to be increased for getting better information from
the $Q_N(T)$ on the atomic inversion. Finally, the results obtained
in this paper are valid also when the field is initially prepared in
the cat states.

\section*{ Acknowledgement}

 The author would like to thank the Abdus Salam International
Centre for Theoretical Physiscs, Strada Costiers, 11 34014 Trieste
Italy for the hospitality and financial support under the system of
associateship, where a part of this work is done.

\end{document}